\documentclass[aps,prl,twocolumn,groupedaddress,floatfix]{revtex4-1}

\usepackage{graphicx}

\usepackage{color}

\usepackage{amsmath}
\usepackage{amssymb}
\usepackage{txfonts}

\newcommand{\dd}[1]{\mathrm{d} #1 \,}
\newcommand{\ddd}[2]{\mathrm{d}^{#1}\! #2 \,}
\newcommand{\eref}[1]{(\ref{#1})}

\newcommand{\reldist}{\Delta}
\newcommand{\bsim}[3]{\makebox[0cm]{\raisebox{ #1 }[0cm][0cm]{\hspace{ #2 } #3 }}}
\usepackage{graphicx}

\begin{document}
\title{Quantum quenches, dynamical transitions and off-equilibrium quantum criticality}

\author{Bruno Sciolla and Giulio Biroli}
\affiliation{Institut de Physique Th\'eorique, CEA/DSM/IPhT-CNRS/URA 2306 CEA-Saclay,
F-91191 Gif-sur-Yvette, France}


\date{\today}

\begin{abstract}

Several mean-field computations have revealed the existence of an out of equilibrium dynamical transition induced by quantum quenching an isolated system starting from its symmetry broken phase. In this work we focus on the quantum $\phi^4$ $N$-component field theory. 
By taking into account dynamical fluctuations at the Hartree-Fock level, corresponding to the leading order of the $1/N$ expansion, we derive the critical properties of the dynamical transition beyond mean-field theory (including at finite temperature). We find diverging time and length-scales, dynamic scaling and aging. Finally, we unveil a relationship with critical coarsening, an off-equilibrium regime that can be induced by quenching from the symmetric toward the symmetry broken phase.

\end{abstract}

\pacs{67.85.-d,03.65.Vf,75.10.-b, 67.85.De}


 \maketitle

\begin{figure}
\includegraphics[width=8.6cm,bb = 41 693 300 792,clip]{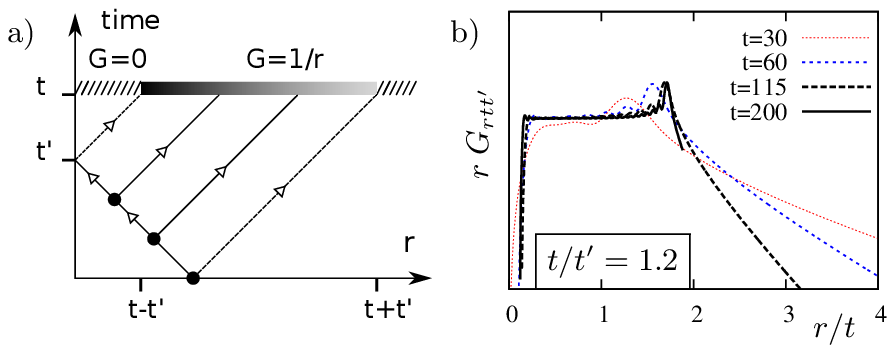}  
    \caption{a) Qualitative interpretation of the correlations in real-space in terms of a common virtual emitter in the past. $G_{rtt'}$ vanishes in the dashed areas, where it is out of causal reach of virtual emitters. b) Rescaled two-times correlation function $r\; G_{rtt'}$ as a function of $r/t$ for $t/t'=1.2$. All data collapse on a step function as $t$ increases, with finite size effects on scale $\Lambda^{-1}$.}
    \label{quasiparticles}
    \label{two_times_correlations}
\end{figure}

Out of equilibrium quantum dynamics of isolated systems is a fundamental research topic which has recently become 
accessible to experimental investigations by trapping ultra-cold atoms in optical lattices \cite{review}. 
After the pioneering work \cite{Greiner}, in which the Mott insulator-superfluid quantum phase transition was observed, 
the field has boomed with a lot of studies, in particular on the so called quantum quenches. These  protocols, consisting in a sudden change of an interaction parameter (for example using Feschbach resonances), bring a system initially in the ground state far from equilibrium. \\
Out of equilibrium quantum dynamics is a very broad field. One of the main fascinating questions is 
whether, and to what extent, there exist universal phenomena generalizing the ones found for equilibrium systems.
The quantum Kibble-Zurek mechanism, describing the production of defects occuring during ramps accross a quantum critical point \cite{Polkovnikov}, is an example of such universal properties. The main topic of this article is another candidate for universal behavior
originally discovered in the Hubbard model~\cite{Werner,Tsuji1,Tsuji2} and later found in a large variety of quantum systems at the mean-field level \cite{Schiro, us,Gambassi}. It consists in a dynamical transition out of equilibrium occurring after a quantum quench. Its main features are that long time averages display a singular behavior 
and the order parameter vanishes when the 
final coupling $U_f$, reached after the quench, approaches a critical value $U_f^d$.\\
Attempts to go beyond mean-field theory in the Hubbard model showed that 
fluctuations play an increasingly important role approaching the transition \cite{Fabrizio2, Fabrizio3}. A full analysis, however, is still lacking. Moreover, 
even though it is recognized that some physical observables are singular at $U_f^d$, the critical nature of the transition remains to be found yet. Actually, it is not known whether there is a diverging correlation length-scale at the transition nor whether 
some kind of critical dynamics scaling takes place. 
In this work we provide answers to these open questions by going beyond mean-field theory and taking into account some dynamical fluctuations.  In order to do that, we shall focus on the $\phi^4$ $N$-components quantum field theory and retain in the self-consistent $1/N$ expansion the leading contributions in the large $N$ limit.
An unexpected and interesting result of our analysis is that the critical out of equilibrium dynamics occurring at the
dynamical transition coincides with the one induced by quenches from the unbroken symmetry phase toward 
the broken symmetry one, a situation similar to the one leading to coarsening dynamics in classical system \cite{Bray}.\\
The model we focus on consists in an $N$ component real scalar field interacting via a quartic term in three dimensions. 
It was studied thoroughly at equilibrium, since, depending on the
value of $N$, it belongs to the same universality class as many physical systems such as 
superfluids and ferromagnets \cite{sachdev}.
The corresponding Lagrangian reads \cite{footnotecte}: \[{\mathcal L}[\phi] = \frac{1}{2} \left ((\partial_t \vec \phi)^2+(\partial_{\bf x} \vec \phi)^2+
r_0 (\vec \phi)^2 
\right)+
\frac{\lambda}{4!N}[(\vec \phi)^2]^2.\]
At equilibrium, this model has a quantum phase transition between a phase with spontaneous symmetry breaking 
in which $\langle \vec{\phi} \rangle$ is aligned along a certain direction for $r_0<r_0^c$ and a paramagnetic phase, $\langle \vec{\phi} \rangle=\vec 0$, for $r_0>r_0^c$. The critical ``mass'' $r_0^c$ is negative, due to the enhancement of the effective mass because of fluctuations. 
It was shown in \cite{Gambassi} that this model displays, at the mean-field level, a dynamical transition due to quantum quenches in the mass $r_0$ (other regimes were previously studied in \cite{Cardy_Sotiriadis}).  
In the following, with the aim of analyzing the effect of fluctuations on the dynamical transition 
we retain in the two-particle irreducible/Baym-Kadanoff expansion of the self-energy
the leading order contribution in $1/N$, which corresponds to the the dynamical Hartree-Fock approximation \cite{Berges}. The initial condition for the dynamics is the ordered ground state before the quench (finite temperature initial conditions will be considered later). Without loss of generality we focus on the case where the average field, $\vec \phi_{t}$, is aligned
along the first component:  $\phi_{t}^n=\delta_{n,1}\phi_t = \langle \hat{\phi}_{{\bf x},t}^1 \rangle$. Note that by symmetry 
the average field remains uniform for $t>0$ and only the diagonal terms $n'=n$ of the connected Keldysh correlation functions, $G^{n n'}_{rtt'} = \langle \{ \hat{\phi}^n_{0,t},\hat{\phi}^{n'}_{{\bf r},t'} \}  \rangle - \phi^n \phi^{n'}$, are nonzero. 
The time-dependent Dyson equations governing the evolution of the system after the quantum quench from 
$r_0^i$ to $r_0^f$ read:
\begin{align}
& \label{field} \partial^2_{t} \phi_{t} = - \left(r_t + \frac{\lambda}{6N} \int_p G^{\varparallel}_{pt t} \right) \phi_{t} = - \frac{\partial V(\phi)}{\partial \phi}\\
& \label{propagator} \partial^2_{t} G^{\bot}_{pt t'} = - \left(p^2 + r_t  \right) G^{\bot}_{pt t'} \\
& \partial^2_{t} G^{\varparallel}_{pt t'} = - \left( p^2 + r_t +  \frac{\lambda}{3N} \phi^2_{t} \right) G^{\varparallel}_{pt t'} \\
& \label{mass} r_t = r_0^f + \frac{\lambda}{6N} \left( \phi_{t}^2 + \frac{1}{2} \int_p G^{\varparallel}_{ptt} + \frac{N-1}{2}\int_p G^\bot_{pt t} \right)
\end{align}
where the parallel index has been used for the $n=1$ Keldysh correlation function and the perpendicular one for all the others (which are equal by symmetry). The initial condition at $t=0$ is given by the value of the field $\phi$ and the equal time ($t=t'=0$) Keldysh correlation function in the ground state corresponding to the value of the mass $r_0^i$. See the EPAPS for more details.
\begin{figure}
    \begin{minipage}[t]{0.49\linewidth}
	\includegraphics[width=\linewidth]{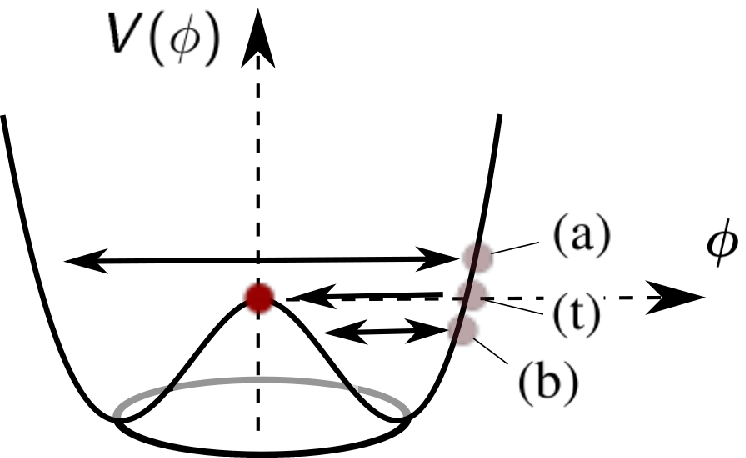}\bsim{2.8cm}{-8.6cm}{a)}
    \end{minipage}
    \begin{minipage}[t]{0.49\linewidth}
	\includegraphics[width=0.9\linewidth,bb = 185 698 293 788,clip]{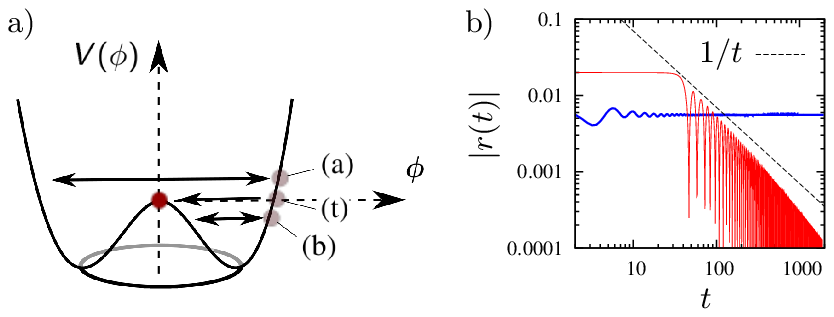}
    \end{minipage}

\caption{\label{mass_evolution}\label{mean_field_dynamical}a) Cartoon of the dynamical transition at the mean-field level. From top to bottom: quench above (a), at (t) and below (b) the dynamical transition. b) $|r_t|$ for a quench within the unbroken symmetry phase (thick blue line) and at the dynamical transition (thin red line). In the second case, $r_t$ decays faster than $1/t$.}
\end{figure}  
Since this problem is not exactly solvable, we integrated numerically the equations for a large value of $N=10^6$ \cite{numerics} (note that the average field 
scales as $\sqrt N$).
Although the dynamics of the field $\phi_t$ and correlations $G^{n n}_{p t t' }$ look superficially similar to a free field evolution, the time dependence of the effective mass $r_t$ has dramatic effects as we shall show. \\
Let us first recall the main result of mean-field theory, which corresponds to neglecting all the feedback of correlations on the dynamics of $\phi_t$ in \eqref{field} \cite{MFphi4}. 
The motion of the field is qualitatively represented in Fig.~\ref{mean_field_dynamical}a, where various quenches with different initial mass $r_0^i$ and with same final mass $r_0^f$ are depicted.
(This means that the potential $V(\phi)$ after the quench is the same. The initial condition instead depends on the value of $\phi$ in the ground state before the quench, {\it i.e.} on $r_0^i$).
Above the transition [case (a)] the field oscillates symmetrically around zero and, consequently, 
is characterized by a zero time average $\overline{\phi}=\lim_{T\rightarrow \infty}(1/T)\int_0^T \dd{t} \phi_t$. 
Below the transition [case (b)] the field oscillates around one minimum of the potential and, hence, is characterized by
 a non-zero $\overline{\phi}$.  In between, at the dynamical transition when $r_0^f = r_0^{f(d)}$ [case (t)] the field relaxes exponentially to zero, {\it i.e.} to the maximum of the potential at $\phi = 0$.
The phenomenology of this mean-field transition is identical to the one found in other mean-field models \cite{Schiro,us}. For example,  the time averaged value of the field has a logarithmic singularity at the dynamical transition: $\overline{\phi} \propto 1 / \ln |\reldist| $, where 
$\reldist$ is the relative distance to the dynamical critical point:
\begin{align}
\label{reldist} \reldist=\left[ r_0^f-r_0^{f(d)} \right] /r_0^{f(d)}
\end{align}
Our goal is to determine the impact of fluctuations at first order in $1/N$ on this scenario.
The numerical analysis of the evolution eqs.(\ref{field}-\ref{mass}) shows
that the system always reaches a steady state at long times \cite{conserved}. This is the first difference with respect to mean-field theory, in which oscillations instead persist 
even at long times. We show in Fig.~\ref{mass_evolution}b, as an example, 
the evolution of the mass for two different quenches: we find that oscillations are damped and $r_t$ 
converges toward an asymptotic value. 
\begin{figure}
\includegraphics[width=8.6cm,bb = 48 688 295 786,clip]{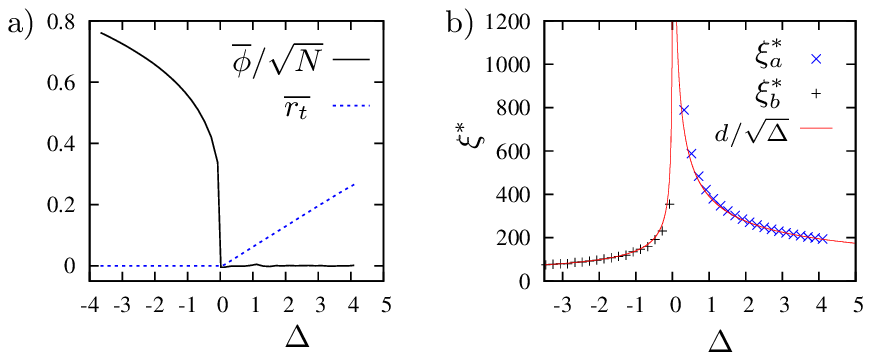}  
    \caption{a) Long time averages, $\overline{\phi}/\sqrt{N}$ and $\overline{r_t}$, as a function of the relative distance to the critical point $\reldist  $ (in \%). b) Critical length $\xi^*$ versus $\Delta$ the distance from the dynamical transition. Notice that despite the different definitions below and above the transition, $\xi^*$ diverges as $d/\sqrt{\Delta}$ on both sides of the transition, with $d_a$ and $d_b$ two different constants.}
    \label{dyn_trans}
	\label{crit_len}
\end{figure}
Similar results are found for the field. 
By studying quenches for several values of the final and initial mass, we find that the dynamical transition 
continues to take place, as it was already mentioned in contexts related to cosmology \cite{Boyanovsky}. In the following, we study its critical features.
Like in mean-field theory, the transition happens for quenches within the regime of broken symmetry: $r_0^i<r_0^c \rightarrow r_0^{f(d)}<r_0^c$ and corresponds to a singularity in the asymptotic value (or equivalently 
the time averaged value) of the field. We show in Fig.~\ref{dyn_trans}a $\overline{\phi}$ and the average mass 
as a function of $\reldist$.
Below the transition, the field relaxes to a nonzero asymptotic value and  $r_t$ vanishes. 
Above the transition, the field relaxes to zero, whereas the mass converges to a positive value.
The critical behavior is different from the mean-field one, since instead of a logarithmic singularity the average field vanishes as  $\overline{\phi} \sim |\reldist|^{1/4}$  approaching the transition from below ($\reldist \rightarrow 0^+$), whereas the asymptotic value of $r_t$ 
vanishes as $\reldist$ for $\reldist \rightarrow 0^-$ \cite{supplementary_mat}. 
After having established the existence of a critical point let us now study its properties, {\it i.e.} focus 
on the physical behavior after quenches right at $\reldist=0$. 
We find that the dynamics is divided in two stages. First, the field relaxes to zero on a timescale $\mathcal T$ smaller than the one characterizing the evolution of $|r_t|$.
In the second stage, $G_{ptt}^\bot$ increases exponentially, as $G_{p00} e^{2\sqrt{-p^2-r_t}t}$, for all momenta below a cutoff $\Lambda^2=|r_{t=0}|$.
This leads to a growth of the effective mass $r_t$, which eventually stabilizes around zero, with a slow, oscillating, power law decay shown in Fig.~\ref{mass_evolution}b.
This in turn stabilizes the growth of $G_{ptt}^\bot$.
At large times, the low momentum modes enter a remarkable \emph{two-times dynamic scaling regime}:                                                                                            
\begin{align}
&G_{ptt'}^\bot\simeq \frac{A}{p^2} \mathcal{F} \left(pt^z,\frac{t}{t'} \right) \label{scaling1}\\
&\mathcal{F}\left(pt,\frac{t}{t'}\right) \sim \cos\left(pt\left(1-\frac{t'}{t}\right) \right) - \cos\left(pt\left(1+\frac{t'}{t}\right)\right) \label{scaling2}
\end{align}
with a dynamical exponent $z=1$ and $A$ a nonuniversal constant. The parallel mode $G^\varparallel$ follows the same scaling law.
\begin{figure}
\includegraphics[width=8.6cm,bb = 62 685 311 786,clip]{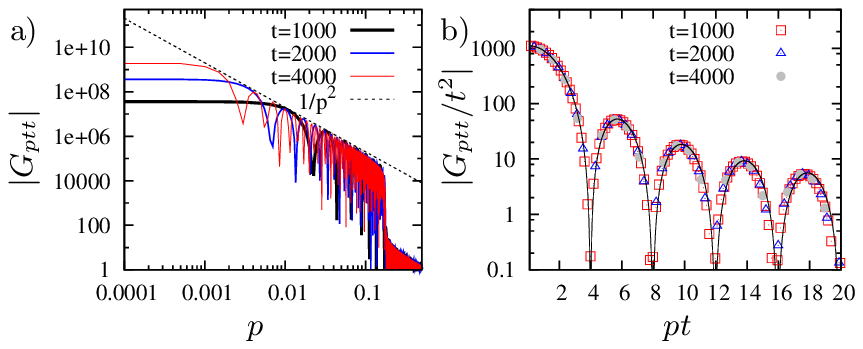}  
    \caption{a) Equal-time correlations $|G_{ptt}|$ as a function of $p$ for $t=\{1000,2000,4000\}$ in a log-log scale. Notice the divergence of correlations below a cutoff scale $p < \Lambda \simeq 0.2$. b) Rescaled equal-time correlations $|G_{ptt}/t^2|$ as a function of $pt$ for the same data, y-axis in log scale. All data collapse on the scaling law \eref{scaling2} drawn in black.}
    \label{correlations_rescaled}
\end{figure}
The real space counterpart of eq. (\ref{scaling2}) reads $G_{rtt'}^\bot \sim \frac{1}{r} \Theta(|r|-(t-t'))\Theta(t+t'-|r|)$.
The existence of the scaling variable $t'/t$ means that the system remains always out of equilibrium: it is not characterized by any intrinsic time-scale besides its age after the quench, a phenomenon called aging \cite{crash}.\\
The scaling \eref{scaling1} and \eref{scaling2} is demonstrated in Fig.~\ref{correlations_rescaled} for equal-time correlations (in Fourier space) and for $t\neq t'$ (in real space) in Fig.~\ref{two_times_correlations}b.
An explanation for the form of the scaling function can be found analyzing quenches in a \emph{free} field theory where the final mass is $r_0^f = 0$. Indeed, by generalizing the result of \cite{Cardy_long} for a sudden quenches in a \emph{free} field theory we find the following expression for the real space two-times correlations in the continuum limit (using the notation $\omega_p^2 = p^2$):
\begin{equation}
G_{r t t'}^\bot = r_0^i \!\! \int \!\! \frac{\ddd{3}p}{(2 \pi)^3} \frac{e^{i \vec{p} \cdot \vec{r}}}{\omega_p^2} \left(  \cos(\omega_p (t-t')) - \cos(\omega_p (t+t')) \right).\nonumber
\end{equation}
This is just the Fourier transform of eq.  \eref{scaling1} and \eref{scaling2}. It's important to realize that, contrary to the free field theory case, now the vanishing mass is \emph{dynamically} generated by interactions. The functional form of the decrease of the mass at long times can be obtained, plugging the dynamical
scaling form of the propagator into \eref{mass}. Calling $\Lambda$ a high momentum physical cutoff, we find:
\begin{eqnarray}
r_t &\simeq& r_0^f+\frac{\lambda}{12} \int_{1/L}^{\Lambda} \frac{\ddd{3}p}{(2\pi)^3} G_{ptt}^\bot\qquad t\gg1 \\
&=&r_0^f+ \int_{1/L}^{\Lambda} \dd{p} \frac{A}{2 \pi^2} (1-\cos(2p t))=-\frac{A}{2 \pi^2} \frac{\sin (2 t \Lambda)}{2t}\nonumber
\end{eqnarray}
where to establish the last identity we have used that the constant contributions cancel since at the transition 
the theory is asymptotically massless. By taking into account sub-leading corrections to the dynamic scaling form of the 
propagator one can show that the mass decays 
even faster than $1/t$ \cite{Boyanovsky,Boyanovsky2}, as indeed we find numerically, see Fig.~\ref{mass_evolution}b. Note that the mapping to a free field theory, valid at large times, is also useful to interpret the form of the
two-times scaling found previously. One has to use that excitations propagate at fixed speed \cite{Cardy_long} and that in the limit of a large number of excitations the fields become classical. 
Then, according to the Huygens-Fresnel principle, plane wave propagation can be interpreted, in three dimensions, in terms of a continuum of virtual emitters. This is illustrated in Fig.~\ref{quasiparticles}a: between the origin and a point at a distance $r$, correlations $G_{rtt'}$ at successive times $t'$ and $t$ are nonzero only provided there is a virtual emitter in the past, susceptible to reach the two points at times $t'$ and $t$ respectively. Notice that this effect includes the usual light-cone effect found in various systems~\cite{Cardy_long,Cheneau}, but that the two-time scaling is really a new feature, due to the critical nature of all effective excitations. Away from the dynamical transition, we still observe the light cone effect but dynamic scaling does not hold any longer.\\
We now analyze how the critical behavior emerge approaching the transition. Note that in this case there are two regimes: First, an out of equilibrium transient that persists for a time-scale $\tau^*_{\textrm{\tiny rel}}$. 
In this regime, corresponding to times $t$ such that $\tau^*_{\textrm{\tiny rel}} \gg t\gg \Lambda^{-1}$, the dynamical scaling 
\eref{scaling1} remains valid (on both sides of the transition) and, hence, the characteristic time-scale 
is the age of the system itself and the characteristic scale for the momentum is the inverse of that.
In the second regime, corresponding to $t\sim \tau^*_{\textrm{\tiny rel}}$, the system reaches a steady state in which the Keldysh correlation function becomes time-translation invariant. The relaxation time-scale to the steady state, $\tau^*_{\textrm{\tiny rel}}$, diverges approaching the dynamical transition. Numerically we found $\tau^*_{\textrm{\tiny rel}}\sim 1/|\Delta|^{1/2}$.\\
In the stationary regime the transverse correlation function becomes time-translation invariant and has a scaling form:
\begin{equation}
G_{ptt'}^\bot=\frac{1}{p^2}F\left(p\xi^*,\frac{t-t'}{\tau^*}\right)
\end{equation}
the low momentum behavior is critical, e.g. $G_{ptt}^\bot\sim 1/p^2$, until values of $p$ of the order of $1/\xi^*$ are reached, accordingly $F(x,y)\rightarrow x^2 f(y)$ for $x\rightarrow 0$. More details on the scaling function can be found in~\cite{supplementary_mat}. Both $\tau^*$ and $\xi^*$ diverge as $ 1/|\Delta|^{1/2}$ approaching the transition \cite{footn}.
The fact that they are characterized by the same critical exponent is in agreement with the unit value of the dynamical
exponent $z$ found previously. The similar divergence of the de-correlation time in the steady state, $\tau^*$,
and the relaxation time toward the steady state, $\tau^*_{\textrm{\tiny rel}}$, can be understood assuming that that there is no intermediate regime. Indeed, if the out of equilibrium evolution stops when 
the typical momentum scale during aging, which is proportional to the time $t$ elapsed after the quench, reaches the steady state value $1/\xi^*$, then one finds $\tau^*_{\textrm{\tiny rel}}\sim1/\xi^*\sim \tau^*$. Note that the asymptotic value of the effective mass $r_t$ is not directly related to $\xi^*$. The latter is determined by studying the low momentum properties of $G_{ptt}^\bot$ whereas the former is relevant only for the dependence in $t-t'$. 
The diverging length is shown in Fig.~\ref{crit_len}b, its divergence is a power law $\xi^*(\Delta) \sim 1/\Delta^{1/2}$, as shown in the supplementary material \cite{supplementary_mat}.\\
A natural question is to what extent starting from the ground
state is important to induce the dynamical transition.
We have addressed this issue, considering quantum quenches from an initial thermal state,
and we find that the dynamical transition remains unaffected, provided the initial state is still in the broken symmetry phase~\cite{supplementary_mat}.
Non-universal features, such as the position of the dynamical transition $r_0^{f(d)}$ instead are different. By increasing the temperature at fixed $r_0^i$ one finds that the value of the 
critical mass approaches $r_0^i$; they become equal when $T$ reaches the value corresponding to the thermal {\it equilibrium} phase transition. For higher temperatures the dynamical transition does not exists any longer.
\begin{figure}
    \includegraphics[width=0.55\linewidth ,bb = 104 687 231 787,clip]{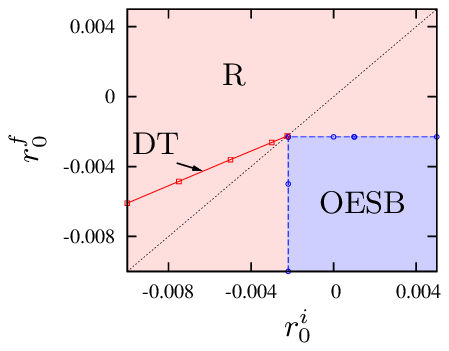}
\caption{Quench phase diagram: long time typical dynamics after a quench $r_0^i \rightarrow r_0^f$ for $\lambda = 1$. DT: Dynamical transition, OESB: Off-equilibrium symmetry breaking, R: Relaxation on large times to a non critical state. Error bars are smaller than item size. The exact position of transition lines depends on non-universal features, such as the interaction strength $\lambda$ and the cutoff $\Lambda$.}
\label{quench_phasediag}
\end{figure}\\
Let us now turn to an apparently unrelated problem: quantum quenches starting from a symmetric ground state, $r_0^i> r_0^c$, toward values of the mass at which the system would be ordered at equilibrium,  $r_0^i<r_0^c$.
This problem has been studied in cosmology and in statistical physics; it is referred to as spinodal decomposition \cite{tachyonic_instability,Bray,Chandran}.
Physically, one expects that the system globally remains in a symmetric state but locally, on length-scales and time-scales that increase with time, it breaks the symmetry. Since the average field remains zero for all times \cite{footnotephi0} and $\overline \phi$ is the only dynamical quantity analyzed at the mean-field level, the latter method is useless to study these quantum quenches. The growth of local order is visible at the level of \emph{correlations}, which requires to go beyond mean-field theory. In the following we briefly present our results obtained at the fleading order in $1/N$.\\
The initial conditions for quenches from the unbroken to the broken symmetry phase correspond to $\phi_t = 0$ (the initial state is symmetric) and negative masses. These are qualitatively similar to those of a quench at the dynamical transition after the time $\mathcal T$ defined above. Indeed, it turns out that the subsequent out of equilibrium dynamics is the same. In particular, the effective mass vanishes asymptotically, and the two-time correlations scale like \eref{scaling2}.
Thus we find that the dynamical transition is characterized by the same critical properties as coarsening dynamics at the leading order in $1/N$.
Note, however, that in usual classical coarsening~\cite{Bray} the equal time propagator is not 
critical since the system is formed by (growing) regions with a definite value of the order parameter. Here, instead, we find a non-equilibrium critical state, akin to the one obtained by quenching to an equilibrium critical point, a phenomenon called critical coarsening~\cite{Calabrese}.
The reason of this discrepancy between quantum and classical cases is unclear: it is the object of ongoing research~\cite{Nowak} and could disappear when $1/N^2$ terms are taken into account. \\
A complete quench phase diagram is shown in Fig.~\ref{quench_phasediag}a, summarizing all possible quenches $r_0^i \rightarrow r_0^f$.
When the initial field is nonzero, $r_0^i < r_0^c$, the system relaxes to a steady state on both sides of the dynamical transition, either to a state of positive field $\overline{\phi}$ or of positive mass $\overline{r_t}$.
The correlations follow the scaling form \eref{scaling1} on the dynamical transition (DT) and in the whole region (OESB) of quenches from the symmetric phase to the broken symmetry phase.\\
In conclusion, by going beyond mean-field theory and taking into account fluctuations at the leading order in $1/N$,
we have shown the existence of an off-equilibrium transition induced by quantum quenches which is characterized by  bona fide critical properties, in particular diverging time and length-scales.
Elucidating the nature of this dynamical transition will be the subject of future works. 
It may be related to either the physics of non-equilibrium thermal fixed points \cite{Berges,Berges2} or of quenches to the thermal critical line \cite{Calabrese}. Recent studies in the Hubbard model favor the former scenario~\cite{Tsuji1,Tsuji2}, whereas the relationship with critical coarsening favours the latter one.  
Clearly, in order to answer this question and generalize our finding to 
systems directly relevant for experiments, it is worth to extend our results to take into account
the next leading order contribution in $1/N$~\cite{Berges2} and to more physical models, such as the Bose Hubbard one. 

{\bf Acknowledgements}\\
We thank L. Cugliandolo, M. Fabrizio, C. Kollath, M. Schir\`o for useful discussion and the ANR FAMOUS for support.


\begin{thebibliography}{}
\bibitem{footnotecte}
For simplicity we absorbed the constant in front of the derivatives by 
redefining the units of time and length.
\bibitem{MFphi4}
The mean-field approximation has been studied within a conformal field theory approach, based on a mapping from the imaginary time evolution to a classical critical film and the analytical continuation to real times \cite{Gambassi}.
\bibitem{conserved}
At the leading order in $1/N$ there is an infinite number of conserved quantities, hence, the system is not expected to thermalize even for ``trivial'' quenches. On the other hand, it was shown that the next leading order contribution allows for thermalization on large times \cite{Berges}.
\bibitem{footn} The parallel modes, unlike the transverse modes, do not follow this reduced scaling because of the presence of the non-zero field.
\bibitem{footnotephi0}
Contrary to the common intuition, a small perturbation of the initial conditions $\phi_{x,t=0} \neq 0$ does not the affect the dynamics very much \cite{tachyonic_instability}, thus we only consider the perfectly symmetric scenario $\phi_{x,t=0}=0$.
\end{thebibliography}
\end{document}